\documentclass[prb,onecolumn,notitlepage,eqsecnum,nofootinbib,floatfix]{revtex4-1}
\usepackage{amsmath}
\usepackage{amssymb}
\usepackage{enumerate}
\usepackage{graphicx}
\usepackage{mathtools}
\usepackage{subcaption}

\makeatletter
\newcommand*{\citenumns}[2][]{%
  \begingroup
  \let\NAT@mbox=\mbox
  \let\@cite\NAT@citenum
  \let\NAT@space\NAT@spacechar
  \let\NAT@super@kern\relax
  \renewcommand\NAT@open{}%
  \renewcommand\NAT@close{}%
  \cite[#1]{#2}%
  \endgroup
}
\makeatother

\begin{document}

\title{Six textbook mistakes in computational physics}
\author{Alexandros Gezerlis}
\affiliation{Department of Physics, University of Guelph, Guelph, Ontario N1G 2W1, Canada}
\author{Martin Williams}
\affiliation{Office of Teaching and Learning \& Department of Physics, University of Guelph, Guelph, Ontario N1G 2W1, Canada}

\date{\today}

\begin{abstract}
This article discusses several erroneous claims which appear in textbooks on numerical methods and computational
physics. 
These are not typos or mistakes an individual author has made, but 
widespread misconceptions.
In an attempt to stop these issues from further propagating, we discuss them here,
along with some background comments. In each case, we also provide a correction, which
is aimed at summarizing material that is known to experts but is frequently mishandled
in the introductory literature. To make the mistakes and corrections easy to understand,
we bring up specific examples drawn from elementary physics and math.
We also take this opportunity to provide pointers to the specialist literature
for readers who wish to delve into a given topic in more detail. 
\end{abstract}

\maketitle

\section{Introduction}

There is no shortage of mistakes in the research literature on computational physics
(or any other subject, for that matter). This is inevitable when the field (or a given tool)
is nascent, since investigators 
have to work ``in the dark.'' Textbooks are another matter: they are designed to condense
the techniques and results that have stood the test of time. It is therefore important to ensure
that mistaken claims that have somehow made it into several textbooks be corrected.
The present work 
studies a set of six misconceptions on themes related to computational science. 
Our intended audience is largely made up of physics instructors who are interested in computational
physics; if they have taught a course
on the subject, they may have even unintentionally contributed toward the propagation
of some of these mistakes.

A selection criterion for the specific mistakes discussed is that
they appear in at least two standard textbooks.
Mistakes by individual textbook authors are sometimes also important, but have a different flavor: we are
here interested in misconceptions which have themselves stood the test of time and are therefore
more pernicious. 
Another criterion we employ is that the mistake should be sufficiently important that
both its statement and its correction can be understood with a minimum of effort. Generally speaking,
this implies that the question it touches upon is of wide import. In other words,
we are not talking about typos or other minor issues that students can figure out on their
own while working through a textbook.
Crucially, the point of the present article is not to criticize authors of respected textbooks.
The aim of our work
is to make sure future generations of students learn these topics right the first time, 
and that is a goal that is better served by elaborating on the mistake and detailing how to fix it,
not by dwelling on the specific wording employed by those who made the mistake.

When possible, we provide the relevant equations and a plot that emphasizes how one
may be led astray. When the specific misconception requires a more detailed argument than we can
provide in this short article, we give pointers to the literature. One of the authors has recently
finished writing a textbook on numerical methods in physics.\cite{Gezerlis}
The points below are handled correctly
in that work, but readers may benefit from seeing alternative viewpoints, so we cite several
related references.

\section{Mistakes and corrections}

Since our goal is not to identify the guilty parties, here we cite, instead,
a superset consisting of many standard textbooks on the subject.\cite{Acton,Ascher,Boudreau,Burden,Dahlquist,Franklin,Garcia,Giordano,
Gilat,Gould,Hamming,Heath,Hoffman,Izaac,Kiusalaas,Klein,Koonin,Landau,Newman,Olver,Pang,Press,Sirca}
(Detailed bibliographic information has been made available to the Editor and Referees of the present manuscript.)
For each of our six themes, we start with some general background material and then provide 
a \textit{mistake}, namely a specific 
quote from the literature exemplifying a misconception. After some discussion of what went wrong, we
conclude with a \textit{correction}, i.e., a concise statement that overcomes the problems in the earlier version.

\subsection{Roundoff error in an operation}

Floating-point numbers are omnipresent: they are used in the overwhelming 
majority of computations in physics (or other technical fields). 
The representation of real numbers on the computer is a mature field, which
has been studied extensively. While in the past there existed several conflicting
conventions, things are much better today: after the adoption of the IEEE 754 
standard for floating-point arithmetic,\cite{IEEE} 
floating-point numbers can be handled dependably and in a portable
manner. 
Most readers will have some idea of what rounding error is; a famous example involves comparing
\texttt{0.1 + 0.2}  with \texttt{0.3}. Using real numbers, $0.1+0.2$ is obviously $0.3$, 
but due to rounding error floating-point numbers behave differently. 
Unfortunately, the treatment of rounding error in introductory textbooks is often misleading, bringing us
to the first misconception:

{\bf Mistake \#1} ``\textit{There is a useful model for approximating how round-off
error accumulates in a calculation involving a large number of steps [...] we view the error in each
step of a calculation as a literal `step' in a \text{random walk}, that is, a walk for which each step is 
in a random direction.}''

Crucially, the claim here is that the rounding error in a mathematical operation (e.g., a subtraction)
can be approximated by assuming that the errors in the two numbers are uncorrelated. 
Such discussions implicitly invoke the \textit{central limit theorem}, assuming that rounding errors
are random and weakly correlated.
A related assumption is that rounding error \textit{accumulates}
as you include more and more numbers in a given calculation, i.e., it arises when you are studying
many numbers but (presumably) doesn't occur when you only manipulate two or three numbers.

Judging by students' feedback in class, it appears that in their perception
this tends to pull in two opposing directions. On the one
hand, students hear the term ``random'' and are usually overcome by 
a sense of impotence: if the error is random, there is
probably nothing one can do about it. That may imply that it's not worth investigating in more detail
exactly what can go wrong (and how to avoid such problems). On the other hand, if rounding error
is random, then one can produce a general theory (as above, employing the central limit theorem),
which suggests that the problem arises only when you carry out many calculations in a row.

\begin{figure}[t]
\centering
\includegraphics[width=0.5\textwidth]{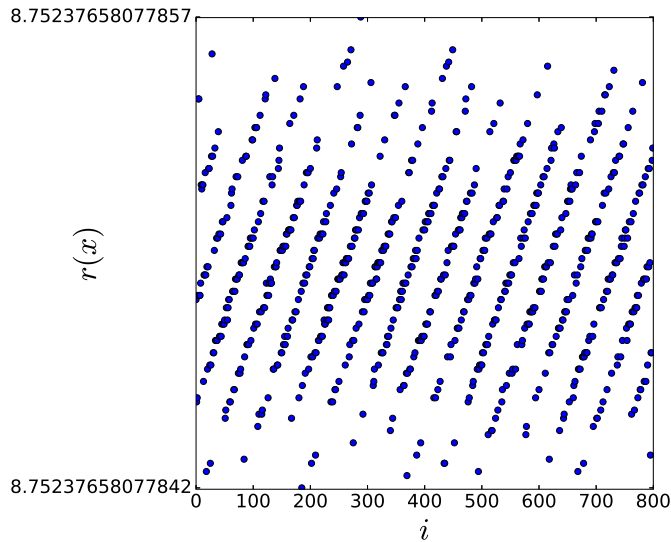}
\caption{Value of rational function from Eq.~(\ref{eq:kahanrational}) for $x = 1.606 + 2^{-52}i$.
The 
typical error is ${\approx}10^{-13}$, but the specific value depends on which $x$
one is investigating.} 
\label{fig:kahan}
\end{figure}

To see why both of these impressions are incorrect, let us introduce a simple example:
using (double-precision) floating-point numbers, 
\texttt{-1.e30 + (1.e30 + 1.)} gives an answer of \texttt{0.}, which is obviously wrong.
If we simply add these numbers in a different order, i.e., 
\texttt{(-1.e30 + 1.e30)  + 1.} we get \texttt{1.}, which is the right answer.
In both scenarios, we were faced with only three numbers: there is no mysterious
round-off error \textit{accumulation} taking place.  
In the first scenario, \texttt{1.e30} and \texttt{1.} are separated by 30 orders of magnitude,
so the answer for that addition is simply \texttt{1.e30}; we would have needed to keep track of
31 significant figures to correctly compute (and store) 
the answer, and double precision only provides us with 15-16 of them.
In the second scenario, the two large numbers cancel each other out, so we are left with the correct final
answer.

One could bring up many more cases along the same lines. 
An example due to W. Kahan,\cite{Kahan} who was instrumental
to the creation of the IEEE 754 standard, is shown in Fig.~\ref{fig:kahan}:
Horner's rule (the standard technique used for accurate polynomial
evaluation) has been used  twice, once for the numerator and once for the denominator of
\begin{equation}
r(x) = \frac{4 x^4 - 59 x^3 +324 x^2- 751x + 622}{x^4 - 14x^3 + 72 x^2 - 151x + 112}~.
\label{eq:kahanrational}
\end{equation}
Notice that this plot is quite zoomed in, so we are essentially seeing the effects of rounding error:
far from being ``noise'' (i.e., uniformly random), this is obeying unmistakable patterns.

These are not hand-picked counter-examples; to the contrary,
it's rather uncommon in practice to see rounding errors combine randomly. 
(Of course, this does not mean that a probabilistic/statistical analysis cannot be a useful tool
in the hands of experts.)
The non-randomness of rounding error 
is discussed eloquently and unambiguously in N. J. Higham's book.\cite{Higham}
To summarize the main point:

{\bf Correction \#1} \textit{Rounding errors are not random and are typically correlated.
When rounding error leads to trouble, instead of thinking about a mysterious rounding-error 
accumulation, you should look for the one or two operations that are to blame. In other
words, there is no substitute for a detailed investigation of where things went wrong.}

\subsection{Determinant value and ill-conditioning}

Linear algebra is employed in most of computational physics. The two main problems that linear
algebra studies can be trivially stated: (a) $\mathbf{A} \mathbf{x} = \mathbf{b}$ is a concise
way of stating a linear system of $n$ equations in $n$ unknowns, and (b) $\mathbf{A} \mathbf{v} = \lambda \mathbf{v}$ is the standard form of the matrix eigenvalue problem. While these two problems are very
easy to write down, entire monographs have been dedicated to their efficient solution; see, e.g., Ref.~\citenumns{Wilkinson}.
Given the emphasis that numerical analysts have placed on the subject, it is not surprising to hear
that state-of-the-art libraries (like BLAS and LAPACK) are robust and dependable. 
However, even a robust library cannot help you if the problem you are trying to solve is pathological.
Thus, it is always good to have an \textit{a priori} estimate of how tricky a given problem is, i.e.,
of knowing what headaches may arise before one launches into a detailed study. 
This is precisely the study of the \textit{conditioning} of a problem;
in what follows, we limit ourselves to the solution of linear systems, $\mathbf{A} \mathbf{x} = \mathbf{b}$.
It is easy to understand
how the following misconception
arose, but that does not make it any less problematic:

{\bf Mistake \#2} ``\textit{For such [ill-conditioned] systems a small change in coefficients will produce large changes 
in the result. In particular, this situation occurs when the determinant for $\mathbf{A}$ is close to zero.}''

On the face of it, this is plausible: a matrix $\mathbf{A}$ which is singular has zero determinant,
$\text{det}(\mathbf{A}) = 0$. It is then tempting (but wrong) to say 
that a near-singular matrix is one that has a near-zero determinant. The first thing that comes to mind
when investigating this tentative criterion is how one could quantify ``close to zero.'' One should be able
to relativize such a determination: a number which encapsulates the entire matrix 
(such as the determinant)
may be large or small in comparison to 
the magnitude of $\mathbf{A}$'s matrix elements. Quantitatively, one introduces the concept
of the matrix norm, which is a number summarizing how large or small the matrix elements are.
For example,
\begin{equation}
\lVert \mathbf{A} \rVert = \sqrt{ \sum_{i=0}^{n-1} \sum_{j=0}^{n-1} \left | A_{ij} \right |^2}
\label{eq:enorm}
\end{equation}
is the Euclidean or Frobenius norm. 
(Other definitions of the matrix norm also exist, but the intuitive point we are making here
applies regardless.)

Since a matrix norm is non-negative, a plausible way to cast the above tentative criterion is as follows:
$|\text{det}(\mathbf{A})| \ll \lVert \mathbf{A} \rVert$, i.e., the absolute value of the matrix 
determinant is much smaller than the matrix norm. This automatically adjusts for the fact
that the matrix elements themselves could be small, in order to see if the determinant 
is even smaller (or not). Let's apply this criterion to an example. For the matrix
\begin{equation}
\mathbf{A} =\begin{pmatrix}
    1\times 10^{-5} & 2\times 10^{-5} \\
    3\times 10^{-5} & 4\times 10^{-5}  \\
\end{pmatrix}  
\label{eq:mat1}
\end{equation}
we find $|\text{det}(\mathbf{A})| = 2 \times 10^{-10}$ and 
$\lVert \mathbf{A} \rVert \approx 5.5 \times 10^{-5}$. Thus, since the criterion
$|\text{det}(\mathbf{A})| \ll \lVert \mathbf{A} \rVert$ is clearly satisfied, you may be 
thinking that this is an ill-conditioned matrix, i.e., a matrix for which many methods will have a hard
time (and unstable methods will simply fail).

At this point, we recall that this matrix $\mathbf{A}$ arises in the study of a linear
system of equations of the form $\mathbf{A} \mathbf{x} = \mathbf{b}$ 
(a $2 \times 2$ system for our example). But this means that we are free to 
multiply our $n=2$ equations by a constant, without changing anything in the statement of the problem.
In other words, multiplying the matrix in Eq.~(\ref{eq:mat1}) with, say, $10^{10}$ should not
lead to a fundamental change. For the matrix
\begin{equation}
\mathbf{B} =\begin{pmatrix}
    1\times 10^{5} & 2\times 10^{5} \\
    3\times 10^{5} & 4\times 10^{5}  \\
\end{pmatrix}  
\label{eq:mat2}
\end{equation}
we find $|\text{det}(\mathbf{B})| = 2 \times 10^{10}$ and 
$\lVert \mathbf{B} \rVert \approx 5.5 \times 10^{5}$. 
In other words, we now find $|\text{det}(\mathbf{B})| \gg \lVert \mathbf{B} \rVert$, 
so this appears to be a (very) well-conditioned matrix. This is the \textit{opposite} conclusion
from that drawn in the previous paragraph!

There is obviously something wrong with the criterion: multiplying all of our equations
with a constant changes the norm, but it has a more dramatic impact on the determinant. However,
such a simple rescaling has nothing to do with how well-conditioned or ill-conditioned our problem 
is. The resolution is to employ matrix perturbation theory and carefully study the effect
a small change of our matrix has on the solution vector, i.e., 
what $\mathbf{x} + \Delta \mathbf{x}$ looks like when you start with $\mathbf{A} + \Delta \mathbf{A}$.
A quick derivation here shows that\cite{Trefethen}
\begin{align}
\frac{\lVert \Delta \mathbf{x} \rVert}{\lVert \mathbf{x} \rVert} \leq \lVert \mathbf{A} \rVert ~\lVert \mathbf{A}^{-1} \rVert ~\frac{\lVert \Delta \mathbf{A} \rVert}{\lVert \mathbf{A} \rVert}~.
\label{eq:mat3}
\end{align}
In words, this relates an error bound on $\lVert \Delta \mathbf{x} \rVert / \lVert \mathbf{x} \rVert$ 
with an error bound on $\lVert \Delta \mathbf{A} \rVert / \lVert \mathbf{A} \rVert$.
The coefficient in front of $\lVert \Delta \mathbf{A} \rVert / \lVert \mathbf{A} \rVert$ tells us
whether or not a small perturbation
will tend to get magnified.
Equation~(\ref{eq:mat3}) naturally leads us to introduce the \textit{condition number}
\begin{equation}
\kappa(\mathbf{A}) = \lVert \mathbf{A} \rVert ~ \lVert \mathbf{A}^{-1} \rVert~.
\label{eq:mat4}
\end{equation}
When $\kappa(\mathbf{A})$ is of order unity we are
dealing with a well-conditioned problem, i.e., a small perturbation is not amplified.
Conversely, when $\kappa(\mathbf{A})$ is large, a perturbation \textit{is} (typically) amplified, 
so our problem is ill-conditioned. Crucially, this condition number involves both the matrix 
$\mathbf{A}$ and its inverse, $\mathbf{A}^{-1}$. (Equally crucially, this 
condition number has nothing to do with the determinant.)
As a result, an overall scaling has no effect:
the condition number for both $\mathbf{A}$ and $\mathbf{B}$, from Eq.~(\ref{eq:mat1})
and Eq.~(\ref{eq:mat2}) respectively, is 15, i.e., not very large. 
The study of problem conditioning, as well as its relationship to the stability of a given method,
is nicely discussed in the book by N. Trefethen and D. Bau.\cite{Trefethen}
To summarize:

{\bf Correction \#2} \textit{To find out whether or not the problem $\mathbf{A} \mathbf{x} = \mathbf{b}$ 
is ill-conditioned,
you should compute the condition number $\kappa(\mathbf{A}) = \lVert \mathbf{A} \rVert ~ \lVert \mathbf{A}^{-1} \rVert$ (and forget about the determinant).}

\subsection{Lagrange interpolation at many points}

In physics, we are very often provided with a \textit{table} of values: we know that these are correct, but it is very costly
(or impossible) to produce more points. Even so, in practical applications we typically need to have access to 
values ``in between'' the table's rows. For example, we may be given a table of points, $(x_j, y_j)$ for $j = 0, 1, \ldots, n-1$ 
(which exactly represent an underlying $f(x)$) but we require the values of the function $f(x)$'s derivative, or integral, and so on.
This is where \textit{interpolation} comes in: one picks a set of $n$ basis functions $\phi_k(x)$ and uses them to produce
a linear form:
\begin{equation}
p(x) = \sum_{k=0}^{n-1} c_k \phi_k(x)~.
\label{eq:approx1}
\end{equation}
The (possibly non-linear) $\phi_k(x)$ are considered known and one attempts to determine the $n$ parameters $c_k$ in such a way
that the interpolating polynomial $p(x)$ goes through our input data points. The first idea that comes to mind
is to use polynomials (or monomials) as basis functions; a time-tested technique that does this goes by the name of ``Lagrange interpolation.''
(This is not least-squares fitting, since we trust
that the input is correct.) Given how fundamental this problem is, you would be justifiably surprised
 to learn that it is 
handled quite
poorly in the literature:

{\bf Mistake \#3} ``\textit{[T]his doesn't work because very high order polynomials tend to have a lot of wiggles in them and can deviate
from the fitted points badly in the intervals between points. It's better in this case to fit many lower-order polynomials such as
quadratics or cubics to smaller sets of adjacent points.}''

As it so happens, Lagrange interpolation has been subject to more than one misconception. Starting with the least pernicious one, 
we quote from the excellent volume by Dahlquist and Bj\"orck (Ref.~\citenumns{Dahlquist}, p. 284): Lagrange's interpolation formula is said to be
``more suitable for deriving theoretical results than for practical computation.'' In other words, Lagrange interpolation is taken to be a tool
for proving theorems, not for actual calculations (F. Acton says something similar in his classic book, 
Ref.~\citenumns{Acton}, pp. 95-96.)
As a result, many volumes on numerical methods
(or on numerical analysis) briefly mention the existence of Lagrange interpolation, before moving on to other approaches (most commonly,
via Newton's divided differences). As explained in N. Trefethen's book on approximation 
theory,\cite{TrefethenApp} 
the criticisms of Lagrange interpolation are unwarranted: using the barycentric interpolation formula allows one to separate the 
construction stage (with a cost of $\mathcal{O}(n^2)$ operations) from the evaluation stage (which costs $\mathcal{O}(n)$ operations).
In other words, once you pick the interpolation nodes, evaluating the interpolating polynomial in between the nodes is very easy.

However, whether or not Lagrange interpolation is efficient is not
the main problem. The quote given in Mistake \#3 above is representative of the overwhelming majority of computational
physics textbooks in questioning the entire premise of using a single polynomial to approximate the behavior of a complicated function.
This is typically called the ``Runge phenomenon'' and is 
sometimes accompanied by a plot, showing a set of data points and an attempt to interpolate through them via a single polynomial:
the polynomial exhibits wild fluctuations, so the authors recommend against employing such an approach.
Thus, in the literature it is customary to read that one should, instead, use cubic (or other) splines to connect a few points at a time, without
attempting to capture the global behavior of the function. As discussed in N. Trefethen's aforementioned text, this is an issue
that is not limited to computational-physics books: similar claims are widespread in the mathematical literature, also. 

The most straightforward way of seeing that these claims are wrong is to ivestigate a specific example. Since interpolation
arises when the function $f(x)$ is difficult to evaluate ``on-the-fly,'' we decide to look at the complete elliptic integral of 
the first kind:
\begin{equation}
K(m) = \int_0^{\pi/2} \frac{1}{\sqrt{1 - m \sin^2\phi}}~d\phi~.
\label{eq:approx2}
\end{equation}
This integral comes up in the study of the classical 
pendulum beyond the small-angle approximation.
To find $K(m)$ for a given $m$, one needs to compute a definite integral; assuming you are not an expert in numerical
integration, you may decide to use Gauss-Legendre quadrature here, an approach with a (deservedly) good reputation.
Unfortunately, estimating the error in computing the definite integral is difficult that way, so you would be reduced to 
computing the integral with an increasing number of points (and checking if the answer changes). In short, this is a good
example of a costly function: if we need to employ $K(m)$ at millions of $m$'s, we would really prefer not to have to compute a definite integral for
each one of them separately.

\begin{figure}[t]
\centering
   \begin{subfigure}{0.49\textwidth} \centering
     \includegraphics[scale=0.39]{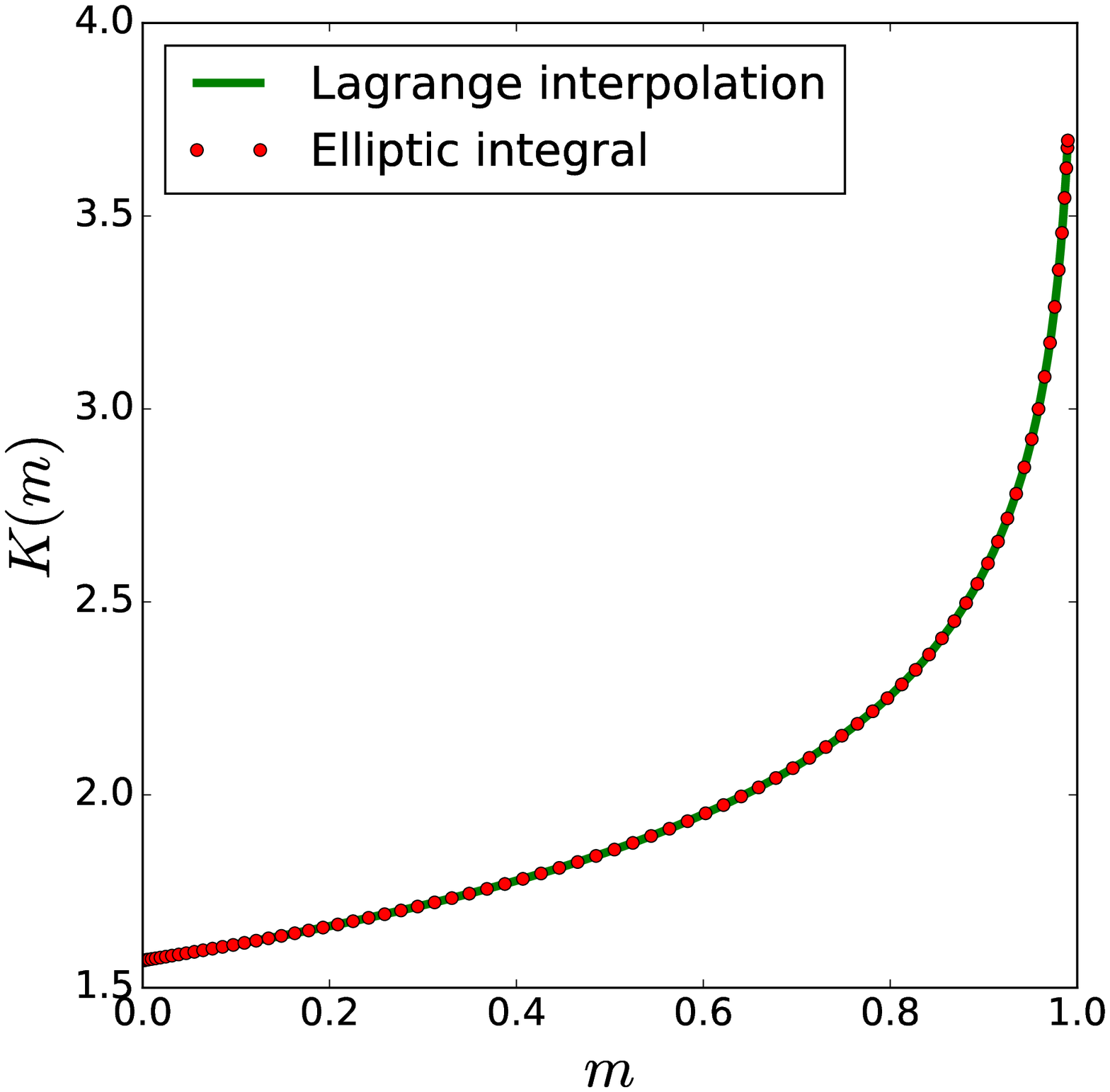}
     \caption{}
   \end{subfigure}
   \begin{subfigure}{0.49\textwidth} \centering
     \includegraphics[scale=0.39]{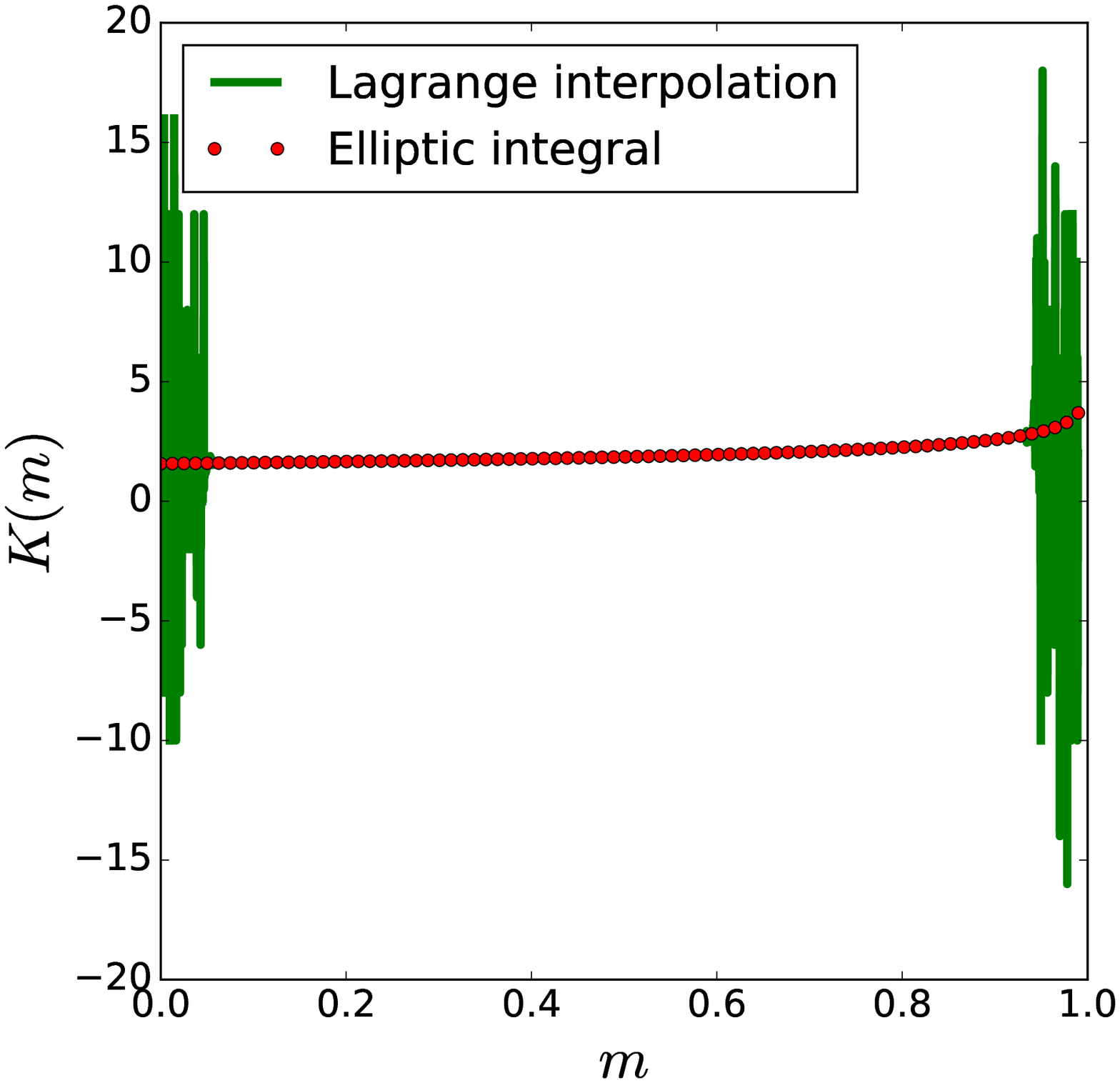}
     \caption{}
   \end{subfigure}
\caption{Data points and interpolated values for the complete elliptic integral of the first kind.
The left panel shows the result of employing Chebyshev nodes, whereas in the right
panel we use equally spaced nodes.} \label{fig:cheby}
\end{figure}

In the spirit of this article, here we don't go into the details of how Lagrange interpolation works. All you need to know in order to grasp
the qualitative point is that we would like to use Eq.~(\ref{eq:approx2}) to compute the value of the 
function at, say, $n=80$ points; we will then produce a polynomial of 79th degree that not only goes through the points, but hopefully also 
captures the underlying function in between. Again, without going through the details of the derivation, we state
the resolution: you would be well advised to pick the $x_j$'s in your table of points
to be \textit{Chebyshev points}, namely the 
extrema of Chebyshev polynomials:
\begin{equation}
x_j = - \cos \left ( \frac{j \pi}{n-1} \right ),~~~~~~~j = 0, 1, \ldots, n-1~.
\label{eq:chebnodes}
\end{equation}
These points lie in the interval $[-1, 1]$; if you are faced with the interval $[a, b]$ you can 
scale appropriately. 
Crucially, these
points cluster near $-1$ and $1$. If you study the error incurred in Lagrange interpolation, you will
see that it strongly depends on this property: as a result, other sets of points which cluster
(e.g., the roots of Legendre polynomials) could also be used here. 

Speaking of Legendre, it's worth noting that most people are comfortable employing 
unequally spaced abscissas when integrating: that is precisely how Gaussian quadrature comes about.
Using equally spaced abscissas for integration leads to wild fluctuations, which is why Newton-Cotes
methods stick to low-degree polynomials for a few points at a time: this is essentially piecewise
interpolation (followed by integration of the interpolant). 
Thus, it should not really come as a surprise that interpolation at \textit{un}equally spaced points does not
lead to any conceptual or numerical problems.

Without further ado, we show the result of employing 80 Chebyshev points for the complete elliptic integral of 
the first kind in the left panel of Fig.~\ref{fig:cheby}. We see that our interpolating polynomial exhibits no wiggles whatsoever, despite
the fact that we employed a high degree intentionally; actually, for this example, we could have 
produced a decent interpolant even with fewer points, but we wanted to emphasize that using a single polynomial
of a high degree is an excellent tool.
The troubles mentioned in the literature arise only if you employ \textit{equidistant} nodes (which don't cluster at the ends
of the interval): you can see this in the right panel of the figure, where 
wild fluctuations appear near 0 and 1.
If you are in control of the node selection, as is often the case in practice, you can simply
choose Chebyshev nodes and avoid any problems. As it so happens, for analytic functions, you get geometric convergence:
adding 10 more points improves the interpolation's quality by an order of magnitude (!). To summarize:

{\bf Correction \#3} \textit{If the interpolation nodes are at your disposal, you should ``bunch'' them near the ends of the interval. 
You can then employ a single polynomial (of potentially high degree)
to capture a function's complicated behavior. For an analytic function, the interpolant
converges geometrically, i.e., the error is a straight line on a semilog scale.}

\subsection{Discrete Fourier transform and trigonometric interpolation}

Trigonometric interpolation is a close relative of the topic we discussed in the previous subsection, namely 
polynomial interpolation. Essentially, the problem is that of interpolation for the case of periodic
functions/signals, so it should not come as a surprise that Lagrange interpolation could also
be used here. Of course, since our new problem is inherently periodic, it is more appropriate to 
employ sines and cosines as the basis functions $\phi_k(x)$ in Eq.~(\ref{eq:approx1}).
Obviously, an equivalent formulation would be to employ complex exponentials, instead.
The latter approach can be framed to take advantage of one of the most successful algorithms of the twentieth century,
the fast Fourier transform (FFT).
The notation and mathematical expressions on the subject can get messy, and as a result several
misleading (if not outright wrong) statements appear in the literature. This brings us to:

{\bf Mistake \#4} ``\textit{[W]e only need consider a finite linear combination
\begin{equation}
q(x) = \frac{1}{n} \sum_{k=0}^{n-1} \tilde{y}_k e^{i k x} 
\label{eq:dft000}
\end{equation}
[\ldots] the function $f(x)$ and the sum $q(x)$ agree on the sample points:
\begin{equation}
f(x_j) = q(x_j),~~~~~~j = 0, \ldots, n-1
\end{equation}
Therefore, $q(x)$ can be viewed as a (complex-valued) interpolating trigonometric polynomial
[\ldots] we will usually discard its imaginary component.}''

(We have tweaked the notation here, to make the following discussion intelligible.)
Before elaborating on why the quoted statement is wrong, let us spend some time seeing whence it arose.
The \textit{discrete Fourier transform} (DFT) starts from $n$ numbers (presumably representing the value
of a function at a set of grid points, i.e., $f(x_j) = y_j$) and uses those to 
produce $n$ Fourier parameters $\tilde{y}_k$:
\begin{equation}
\tilde{y}_k = \sum_{j=0}^{n-1} y_j e^{-2 \pi i k j/n},~~~~~~~k  = 0, 1 \ldots, n-1~.
\label{eq:dft12}
\end{equation}
The argument can be followed in the opposite direction, giving rise to the \textit{inverse} discrete
Fourier transform:
\begin{equation}
y_j = \frac{1}{n} \sum_{k=0}^{n-1} \tilde{y}_k e^{2 \pi ik j/n},~~~~~~~j  = 0, 1 \ldots, n-1~,
\label{eq:dft14}
\end{equation}
which starts from the Fourier parameters $\tilde{y}_k$ and produces the function values $y_j$.
Observe that both the direct and the inverse DFT refer to a grid of values: there is no continuous $x$
variable anywhere in sight. There are only discrete indices, $j$ and $k$; as a matter of fact,
we are here employing the \textit{shifted} or \textit{standard order} of the DFT, whereby
both the $j$ and the $k$ indices take on the same values (from 0 to $n-1$).

It should now be easy to see what gave rise to the expression in Eq.~(\ref{eq:dft000}):
this looks like a generalization of Eq.~(\ref{eq:dft14}) for any $x$ value. In other words,
this ``interpolating'' function $q(x)$ certainly does go through the grid values, i.e., $q(x_j) = y_j$,
as it should. The question that one should consider, however, is whether or not it is legitimate 
to consider $q(x)$ as an \textit{interpolating} polynomial, i.e., whether it is effective in capturing
the underlying behavior even outside the grid points. To answer this question, we consider 
the Mathieu equation, which appears in the study of 
string vibrations; since our example in the previous subsection corresponded to the classical
pendulum, it is fitting that the Mathieu equation also shows up in the 
case of the quantum pendulum:
\begin{equation}
f''(x) = \left ( - s + 2a\cos 2x \right)f(x),~~~~~~~f(0) = f(2\pi)~.
\label{eq:evp1}
\end{equation}
Mathematically, this is an eigenvalue problem, i.e., a messier version of a boundary-value problem. This
would need to be solved numerically: that is not the focus of our discussion here, though. 
(One could elucidate the relevant concepts via simpler examples, but in this section and the previous
one we have opted for ``realistic'' scenarios; interpolation becomes necessary precisely when
one doesn't want to repeat costly computations more times than is strictly necessary.)
Just take it 
for granted that you have managed to solve this equation (for a given value of $a$) at a small number
of points, say $n=8$, but you would like to interpolate between those points 
so that you can access the underlying function at millions of distinct $x$'s. 

\begin{figure}[t]
\centering
   \begin{subfigure}{0.49\textwidth} \centering
     \includegraphics[scale=0.39]{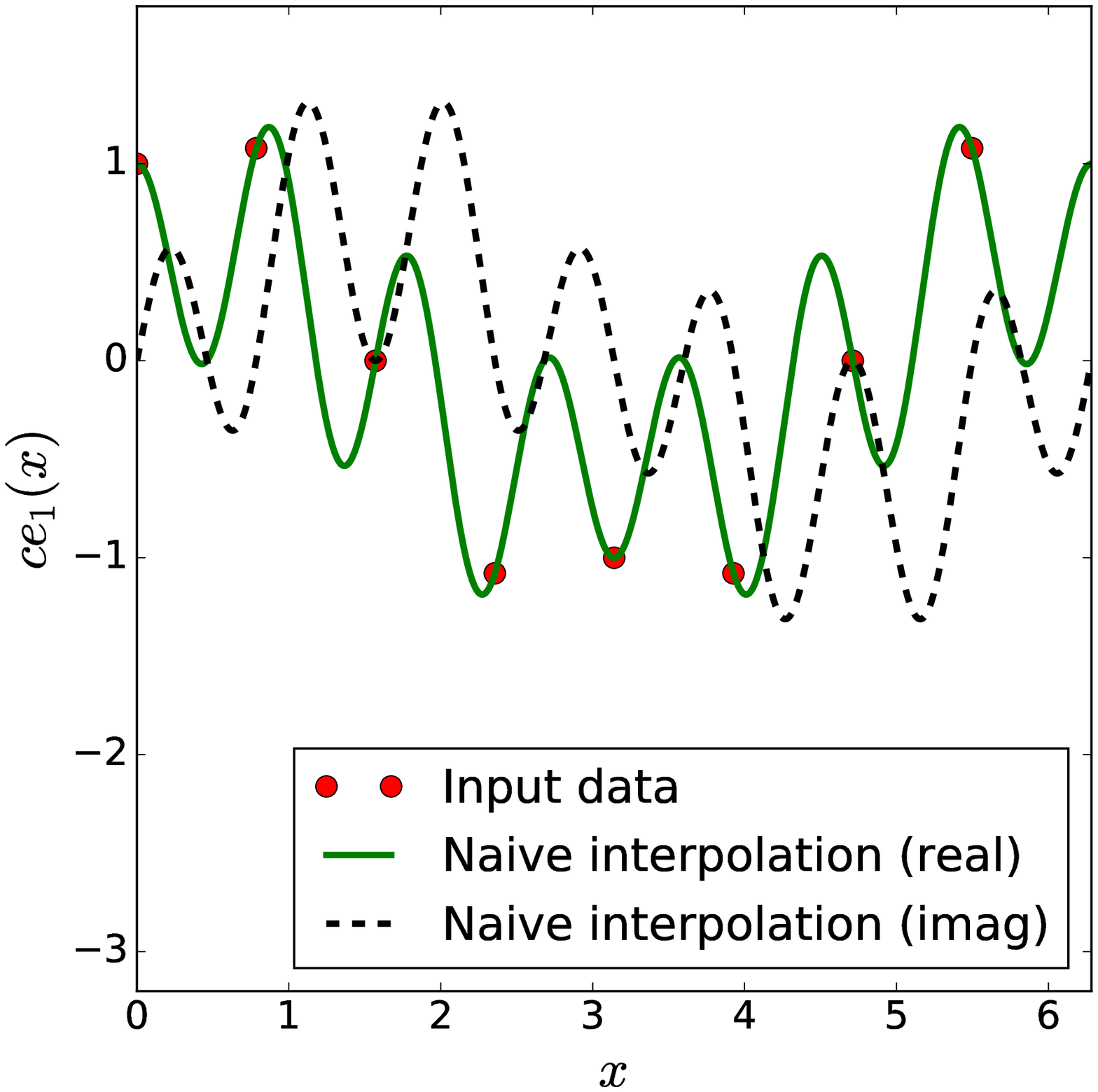}
     \caption{}
   \end{subfigure}
   \begin{subfigure}{0.49\textwidth} \centering
     \includegraphics[scale=0.39]{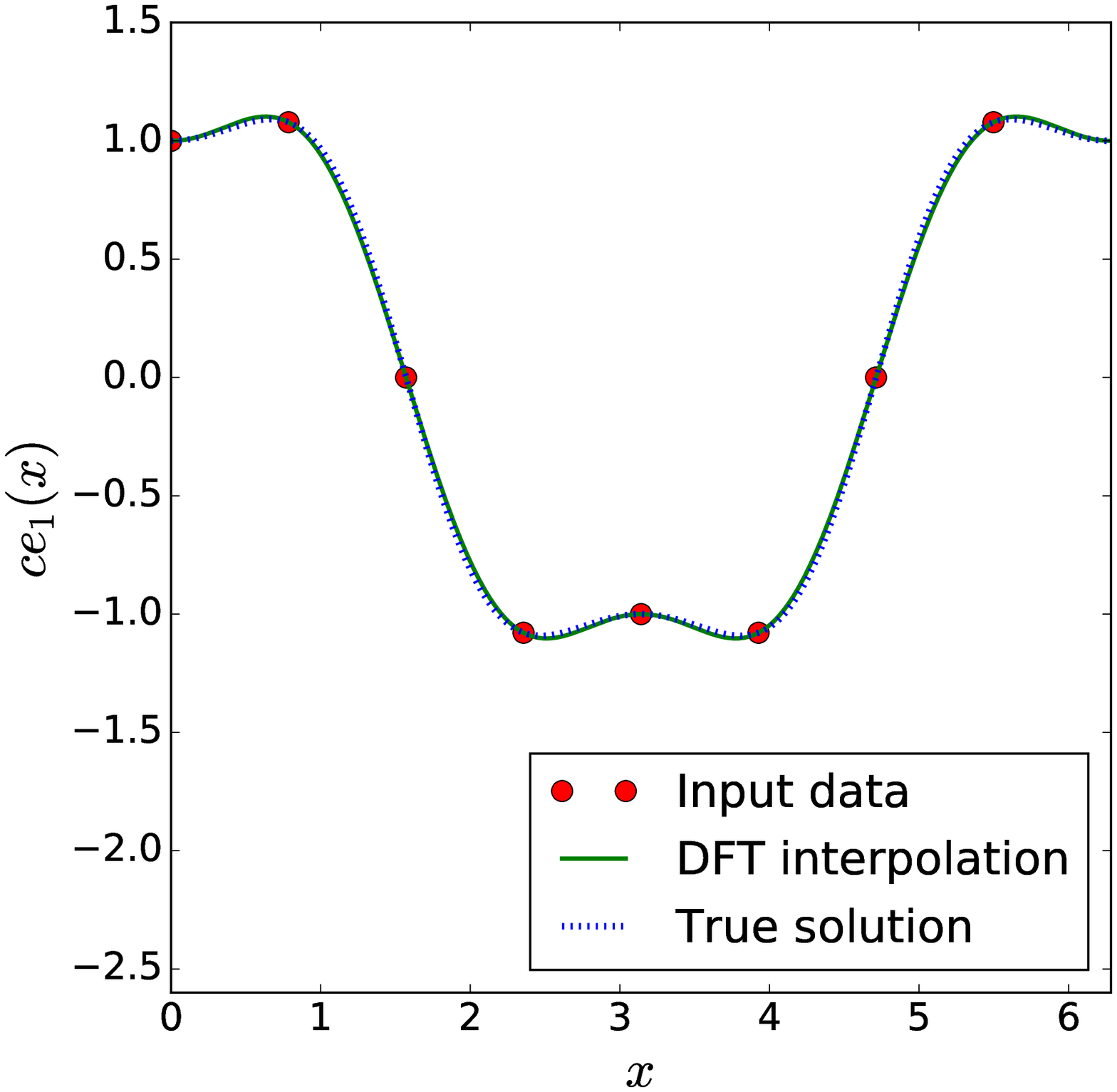}
     \caption{}
   \end{subfigure}
\caption{Incorrect (left) and correct (right) way to carry out discrete Fourier interpolation.} \label{fig:fftinterp}
\end{figure}

In the left panel of Fig.~\ref{fig:fftinterp} we show our input data points $y_j$; they correctly represent  
the second even Mathieu function, denoted $ce_1(x)$, i..e., at those $x$ values we know
that we get an accurate solution, $ce_1(x_j) = y_j$. (The following argument 
would hold just as well if we were discussing a different solution.)
In the same plot, we also show the real and imaginary parts of $q(x)$. 
You can immediately see that $q(x)$ has a sizable imaginary part (off the grid points)
and several fluctuations in between the data points; these fluctuations do not appear to be supported
by the data. In short, the fact that $q(x)$ goes through the points (i.e., $q(x_j) = y_j$) is not enough:
you could come up with another function which oscillates even faster and also goes through the 
points. Neither of these functions can be taken as ``the'' interpolating trigonometric polynomial.

To drive the point home, the right panel of Fig.~\ref{fig:fftinterp} shows the exact solution
$ce_1(x)$, along with the same 8 data points as in the left panel. As we had suspected, the true
solution does not exhibit strong oscillations in between the points.  In the right panel we also
took the opportunity to show the result of employing ``DFT interpolation'': this does not
refer to Eq.~(\ref{eq:dft000}) but to the following, undoubtedly more complicated, equation:
\begin{equation}
p(x) = \frac{1}{n} \sum_{k=0}^{m-1} \tilde{y}_k e^{ikx} + \frac{1}{n} \tilde{y}_{m} \cos m x
+ \frac{1}{n} \sum_{k=m+1}^{n-1} \tilde{y}_k e^{i(k-n)x}~,
\label{eq:fftinterp2}
\end{equation}
where $m = n/2$ and $n$ is even.
Note that this is a general expression for the case of trigonometric interpolation when you have
access to the Fourier parameters $\tilde{y}_k$, i.e., applying this to the Mathieu
equation is only a specific illustration of more general principles.
The details of how Eq.~(\ref{eq:fftinterp2}) came to be do not interest us here (for more, 
see Ref.~\citenumns{Gezerlis}, pp. 358-359).
The important takeaway is that $p(x)$ manages to capture the complicated behavior
of $ce_1(x)$, even though we provided it with only $n=8$ input data points. Even better, it
accomplishes this task without ever giving rise to an imaginary part: since our input
data points were real, any interpolating function should also be real. If you were feeling
uncomfortable about dropping the imaginary part of $q(x)$ ``by hand,'' as the 
quote above instructed you to do,
you can relax now.
To summarize:

{\bf Correction \#4} \textit{The discrete Fourier transform takes you from the $n$ values
$y_j$ to the $n$ values $\tilde{y}_k$. The inverse DFT goes the other way around, but still
only takes you from $n$ values to $n$ values. If you wish to interpolate between these $n$ 
data points, you will need to carefully examine what constitutes trigonometric interpolation, instead
of blindly generalizing the definition of the inverse discrete Fourier transform.
}


\subsection{Acceptance rate in the Metropolis algorithm}

Quadrature techniques like Gauss-Legendre (mentioned above) get into serious trouble when the dimensionality
of the problem grows. Essentially the only approach that works for such problems is 
\textit{stochastic integration}, also known as Monte Carlo integration.
Multidimensional Monte Carlo integration almost always employs the Metropolis 
algorithm, which is (like the FFT) one of the most important algorithms developed in the twentieth century.
More details will be provided below, but a crucial aspect of this algorithm is that it involves
\textit{proposed} steps in a multidimensional random walk: these steps may be either accepted
or rejected. This brings us to:

{\bf Mistake \#5} ``\textit{The actual value of the step size $h$ is determined from the desired accepting rate (the ratio of the accepted to the attempted steps). A large $h$ will result in a small accepting rate. In practice,
$h$ is commonly chosen so that the accepting rate of moves is around 50\%.}''

This seems harmless enough: in many textbooks it is even explicitly called a ``rule of thumb,'' implying
that it's not really based on theory. The problem with this recommendation, as experience 
mentoring undergraduate and beginning graduate students has shown, is that newcomers treat 
this value of 50\% as a goal, i.e., consider a computation as having succeeded if the acceptance rate
is near 50\% (and, therefore, as having failed if the value of this quantity is, say, 30\%). 
As a consequence, students may be overconfident about the trustworthiness of their results or, even
worse, may spend many hours of work trying to tune something which does not add to their understanding
of the physical processes at play. Their time would be better spent exploring detailed features
of their Monte Carlo run.

Let's elucidate what's at stake here. The goal of the Metropolis algorithm is 
to sample from a (known) $d$-dimensional distribution $w(\mathbf{X})$. It accomplishes
this by starting from a location $\mathbf{X}_{i-1}$ and proposing a step as follows:
\begin{equation}
\mathbf{Y}_i = \mathbf{X}_{i-1} + \theta  \times \mathbf{U}_i~.
\label{eq:metro1}
\end{equation}
Here $\mathbf{U}_i$ is a $d$-dimensional set of random numbers from $-1$ to $+1$,
$\theta$ is the ``step size'' in this $d$-dimensional space, and $\mathbf{Y}_i$ is 
the proposed configuration. The heart of the Metropolis algorithm consists of 
computing the ratio $w(\mathbf{Y}_i)/w(\mathbf{X}_{i-1})$ and using that to determine
whether to accept or reject the proposed step.
The question naturally arises of how to pick the value of the step size, $\theta$.
The aforementioned rule of thumb is to pick $\theta$ such that roughly half of the total steps
are accepted (and therefore half are rejected); the ratio of accepted steps over
total proposed steps is known as the \textit{acceptance rate}. 
The thinking behind this is as follows: if you make tiny steps, then you will get a large
acceptance rate but you end up not moving very far from where you started, so your sampling is not very
efficient. If you make very large steps, then your acceptance rate will be small, but that means
that you are wasting your time computing steps which are mostly not accepted
and therefore don't take you anywhere.
The idea is that you should pick the step size somewhere between these two extreme options.

While this qualitative argument is valid, it doesn't actually justify using an acceptance rate of 50\%.
Crucially, the configurations referred to in the previous paragraph are \textit{correlated} since that's 
how they are produced: you start with one configuration and use it to find the next one
(this is known as a \textit{Markov chain}).
The question then arises how to go from this Markov chain, consisting of correlated samples, 
to a set of \textit{un}correlated samples, which have lost all memory of earlier states.
In other words, one must examine the \textit{autocorrelation time} $T$, after which the memory/correlation
gets washed away. If your full simulation runs for much longer than $T$, then you can view it as being made
up of several uncorrelated samples. 
In equation form, we have:
\begin{equation}
T = 1 + 2 \sum_{k=0}^{N-1} C(k)~,
\label{eq:metro2}
\end{equation}
where $C(k)$ is known as the ``autocorrelation function,'' defined as follows:
\begin{equation}
C(k) = \frac{1}{(N - k)(\overline{f^2} - \overline{f}^2)} \sum_{i=0}^{N - k -1} f(X_i) f(X_{i+k})~.
\label{eq:mc3_ag}
\end{equation}
Qualitatively, the autocorrelation function can help us tell how the variance of the mean for statistically
independent values (i.e., the $\overline{f^2} - \overline{f}^2$ in the denominator) gets amplified
due to the fact that the samples \textit{are} correlated; the numerator is known as the 
``autocovariance.'' The autocorrelation function measures the correlation between
configurations which are $k$ steps apart; as shown in the first equation, to get the autocorrelation 
time we sum up all the possibilities. For a given problem, a small $T$ means that the Markov
chain decorrelates fast, i.e., the entire calculation's error bar gets smaller more quickly; that leads
to an overall more efficient Monte Carlo run.

\begin{figure}[t]
\centering
   \begin{subfigure}{0.49\textwidth} \centering
     \includegraphics[scale=0.39]{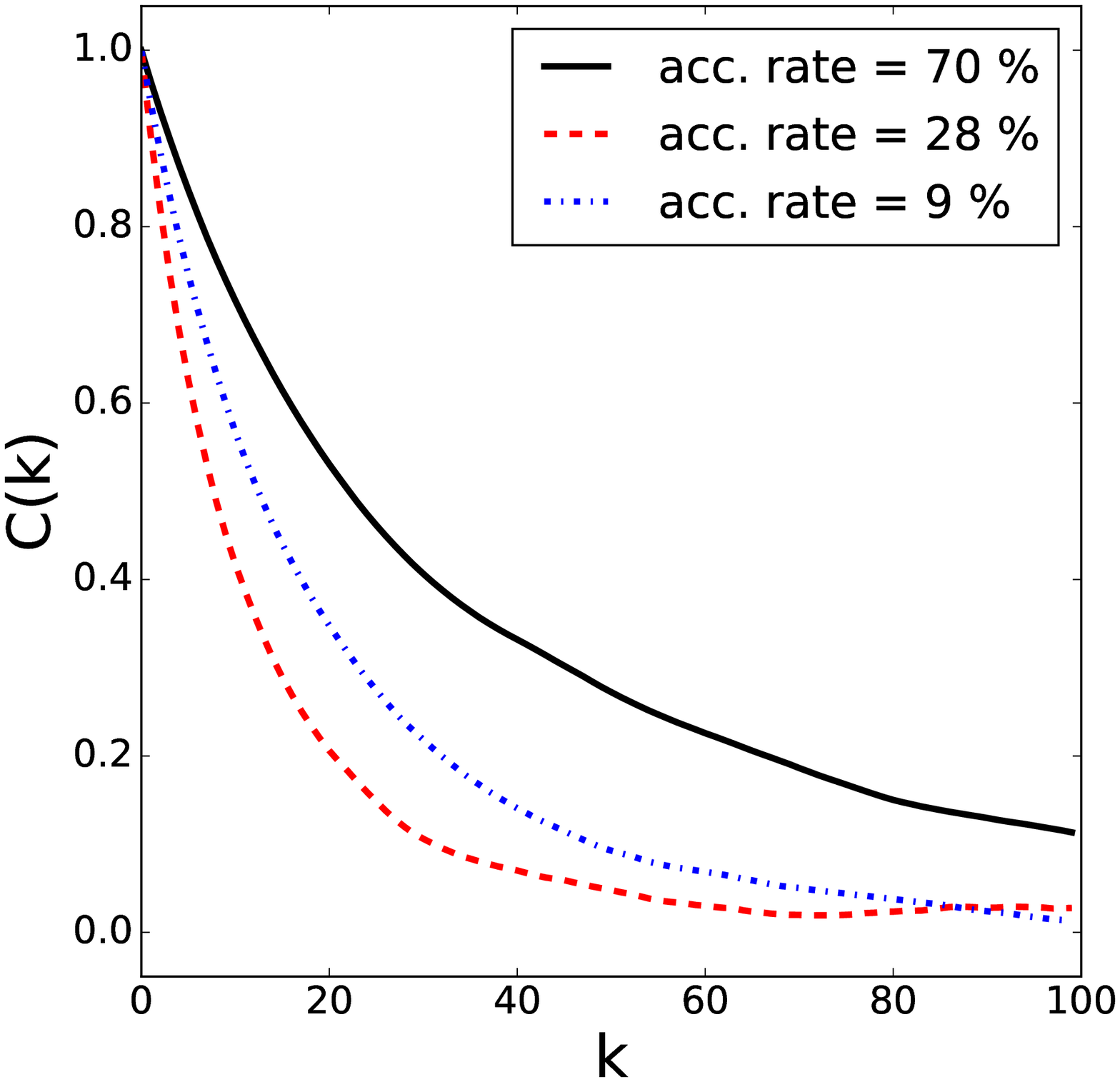}
     \caption{}
   \end{subfigure}
   \begin{subfigure}{0.49\textwidth} \centering
     \includegraphics[scale=0.39]{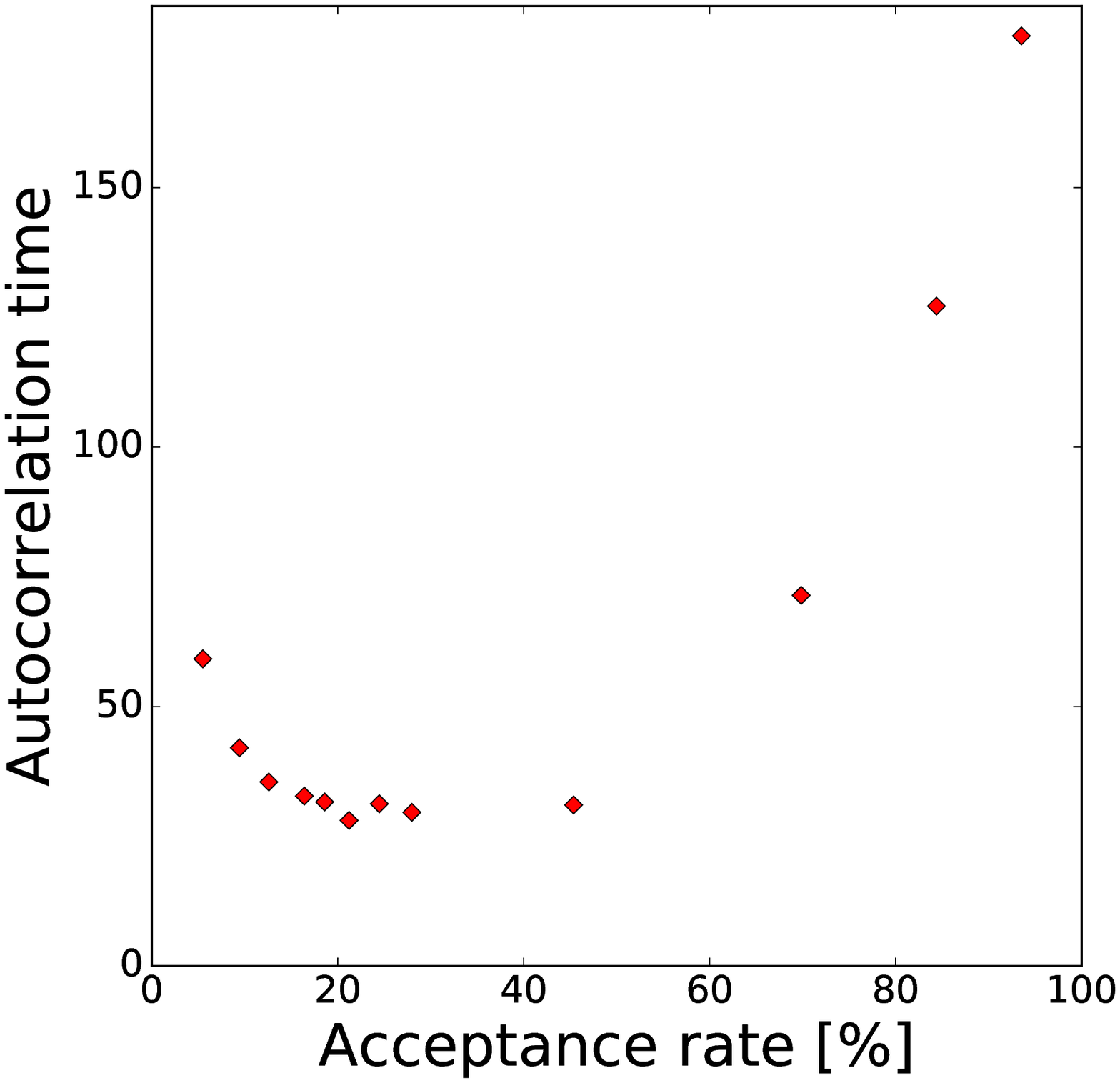}
     \caption{}
   \end{subfigure}
\caption{Autocorrelation function (left) and autocorrelation time (right) in variational Monte Carlo.} \label{fig:efficiency}
\end{figure}

After being introduced to Newton-Cotes or Gaussian quadrature, 
students often wonder why they need 
to learn about Monte Carlo and the Metropolis algorithm.
Put differently, the above discussion may feel too abstract if you've never encountered these concepts before. 
To make things concrete, we will discuss a many-dimensional integral (where Monte Carlo becomes necessary);
as a side-benefit, our specific example is closely related to research problems and may therefore help
spark the curiosity of undergraduate students.
Take four particles (of mass $m$) in three spatial dimensions, in the presence of a harmonic oscillator potential, also interacting with each other in the context of quantum mechanics:
\begin{equation}
\hat{H} = -\frac{\hbar^2}{2m} \sum_{j = 0}^{3}\nabla_j^2 + \frac{1}{2} m ~\sum_{j = 0}^{3}
\left \{ \omega_x^2 [(\mathbf{r}_j)_x]^2  + \omega_y^2 [(\mathbf{r}_j)_y]^2 
+ \omega_z^2 [(\mathbf{r}_j)_z]^2  \right \}
+ g \sum_{\mathclap{\substack{ j, k = 0\\
j<k}}}^{3} \exp\left [- (\mathbf{r}_j - \mathbf{r}_k)^2 \right ]~.
\label{eq:vmc1}
\end{equation}
The Laplacian $\nabla_j^2$ takes the second derivative with respect to the position $\mathbf{r}_j$. 
The frequencies $\omega_x$, $\omega_y$, and $\omega_z$ control the strength of the 
oscillator in each direction and $g$ controls the strength of the interparticle interaction; observe that we are careful not to double-count.
The technique known as \textit{variational Monte Carlo} (VMC) makes use of 
the Rayleigh-Ritz principle to produce an upper bound estimate on the ground-state energy
of the many-particle system. To keep things interesting, we study the case of four interacting
fermions, which have to obey the Pauli exclusion principle; thus, the single-particle states need
to be placed inside a Slater determinant. For this problem, 
the Metropolis algorithm is employed to carry out a 12-dimensional integral.

The left panel of Fig.~\ref{fig:efficiency} shows the autocorrelation function as per 
Eq.~(\ref{eq:mc3_ag}). It's easy to see that the autocorrelation function
starts at 1 and then decays: how fast it does that will determine how quickly the memory of past
configurations is washed away. To produce the three curves shown, we have picked specific
values of the step size $\theta$ in Eq.~(\ref{eq:metro1}) and made a note of the acceptance 
rate that corresponds to each VMC run. This exercise can be repeated for several other values 
of $\theta$. If one also sums up the values of $C(k)$ as per Eq.~(\ref{eq:metro2}), then 
one can produce a plot of
the autocorrelation time $T$ vs the acceptance rate. The result is shown in the right panel of 
Fig.~\ref{fig:efficiency}: since the optimal acceptance rate is that which minimizes the autocorrelation
time, for this case one would pick an acceptance rate of 20\%. 
What we did here is discussed eloquently and unambiguously in the 
book by Martin, Reining, and Ceperley:\cite{Ceperley} ``\textit{There is a rule-of-thumb
that the acceptance should be 50\%, but the correct criterion is to maximize the efficiency.}''

Two further remarks: first, while an acceptance rate of 20\% leads to the 
minimum $T$ in our figure, the behavior is pretty flat near that value. This means that 
choosing a $\theta$ which corresponds to an acceptance rate of, say, 15\% to 50\% should
still be efficient enough for practical purposes. In other words, a student doesn't have to 
waste time carefully tuning the acceptance rate. Second, the calculation carrried out
here is quite costly: plugging Eq.~(\ref{eq:mc3_ag}) into Eq.~(\ref{eq:metro2}) leads to a double
summation. In practical calculations, this is not so easy to do each time one turns to a new physical
system/algorithm. Fortunately, statisticians have proven under quite general conditions\cite{Roberts}
that the optimal acceptance rate is 23.4\% for the many-dimensional case and
44\% for the one-dimensional case. This is consistent with our detailed investigation in the 
right panel of Fig.~\ref{fig:efficiency}. In other words,  our earlier conclusion that 
an acceptance rate from 15\% to 50\% should be efficient enough is of wider applicability. 
In short:

{\bf Correction \#5} \textit{To properly determine what acceptance rate is optimal, you should
try to minimize the autocorrelation time. If you don't want to do that, you will typically 
be OK if your acceptance rate is from 15\% to 50\%, so you should not spend time trying
to tune it to be around 50\%.
}

\subsection{Global error when solving initial-value problems}

Differential equations pop up everywhere in physics, see for example Newton's second law or the Schr\"odinger
equation. 
The study of dynamics is one of the most important (if not \textit{the} 
most important) aspects of physical theory;
mathematically, this gives rise to an \textit{initial-value problem} (IVP):
starting from a given point, one tries to determine the solution(s) at later points.
This is known as the ``numerical integration'' of differential equations, which shares similarities
with numerical quadrature, i.e., the computation of definite integrals, but is also conceptually distinct.
There are several ways of grouping the different techniques designed to tackle
ordinary differential equations: one can 
distinguish between  single-step methods (e.g., Euler's method or Runge-Kutta methods)
and multistep methods (e.g., Adams-Bashforth or Adams-Moulton methods). 
In what follows, we will specialize to the case of single-step methods. In introductory
discussions of such methods, one often hears the terms \textit{local error}
and \textit{global error} being mentioned. Statements like the following
one are commonly made; to make matters worse, such an argument and result is 
often said to be ``intuitive'':

{\bf Mistake \#6} ``\textit{For example, say n steps of the Euler method have been iterated -- we know that
\begin{equation}
x_{n-1} = x_0 + nh
\end{equation}
or rearranging this
\begin{equation}
n = \frac{x_{n-1} - x_0}{h}
\end{equation}
Thus, the total global error must be bounded by
\begin{equation}
n \times \text{step error} = \frac{x_{n-1} - x_0}{h} \times \frac{1}{2} h^2 y''(x) = \mathcal{O}(h)
\end{equation}
As the global error scales linearly with h, the forward Euler method is a first-order algorithm.}''

(We have corrected a typo in the last equation.)
Let us try to unpack this a little. Generally speaking, we are interested in solving an IVP:
\begin{equation}
y'(x) = f\left (x,y(x) \right ),~~~~~~~y(a) = c~.
\label{eq:ivp_again}
\end{equation}
Here $y' = dy/dx$ and $f(x, y)$ is known. We want to solve for $y(x)$, with $x$ going
from $a$ to $b$. 
It is standard (though certainly not mandatory) to employ a discretization scheme 
with equally spaced abscissas:
$x_j = a + j h$
where $j=0, 1, \ldots, n-1$. The step size $h$ is given by
$h = (b-a)/(n-1)$.
At a given $x_j$, we call the exact solution of our differential equation $y(x_j)$ and
use the symbol $y_j$ for the approximate solution resulting from one of our single-step
methods (e.g., Euler's method).
Qualitatively, the \textit{local error} tells you how how close to or how far away from
a given $y(x_j)$ the corresponding $y_j$ is, when you take a single step. 
Similarly, the \textit{global error}
tells you how poorly you have done overall, i.e., how close $y_{n-1}$ is to $y(b)$.
We will come back to the distinction between these two concepts below.

For now, let us go over a likely explanation of why the quoted misconception arose in the first
place. To do so, we first make a quick digression into Newton-Cotes methods, which
use equally spaced abscissas
to compute a definite integral. These approaches slice up the total area 
under a curve
into little rectangles, trapezoids, and so on.
For concreteness, let us stick to the simplest such approach, namely the rectangle rule. As you can
see by taking a simple Taylor series and integrating it from $x_j$ to $x_{j+1}$,
the absolute error incurred by the rectangle rule in a single panel is 
${ \cal E}_{j} = h^2 f'(\xi_j)/2$
where $\xi_j$ is a point between $x_{j}$ and $x_{j+1}$.
Crucially, the error budget in each panel is totally independent from the error budget in each other
panel. Thus, it is straightforward to compute the error in the \textit{composite} rectangle rule
simply as the sum of the one-panel errors; this leads to ${\cal E} = h  (b-a) f'(\xi)/2$, 
where $\xi$ lies somewhere between $a$ and $b$.
Overall,  our result is that the leading error 
in the composite rule is $\mathcal{O}(h)$. This is to be compared with 
the one-panel error, which we saw above was $\mathcal{O}(h^2)$.
So far, things behave exactly as our quoted statement said they would.

\begin{figure}[t]
\centering
\includegraphics[width=0.4\textwidth]{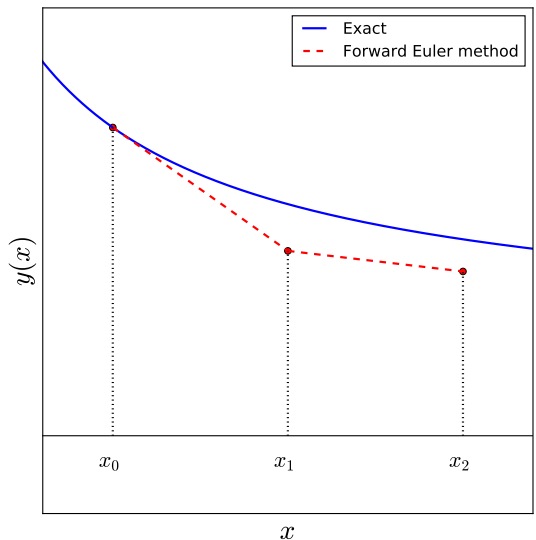}
\caption{Carrying out two integration steps using the forward Euler method.} 
\label{fig:euler}
\end{figure}

Of course, our quote in Mistake \#6 was talking about the solution of initial-value problems,
not numerical quadrature, so let's get back to the numerical integration of differential equations.
Every textbook on this subject introduces
the simplest possible technique, the (forward) Euler method:
\begin{align}
&y_{j+1} = y_{j} + h f(x_j, y_j),~~~~~~~j = 0, 1, \ldots, n-2  \nonumber \\
&\qquad \qquad \qquad \qquad y_0 = c~.
\label{eq:foreu2b}
\end{align}
It's easy to understand the geometrical interpretation of this method:
the right-hand side of Eq.~(\ref{eq:foreu2b}) has a term proportional to $f(x_j, y_j)$;
we know from Eq.~(\ref{eq:ivp_again}) that $f(x_j, y_j)$ is an approximation to
the slope of the tangent to the true solution at $x_j$.
This is illustrated in Fig.~\ref{fig:euler}; for now, focus only on the first
step, for which we have assumed that $y(x_0) = y_0$ since that's where we start.

As you may recall,  in the rectangle rule the one-panel error ${ \cal E}_{j}$ was arrived at by integrating over one
panel. The analogous quantity for initial-value problems, the local truncation error, can be 
defined as follows:
\begin{equation}
t_j = y(x_{j+1}) - y(x_{j}) - h f(x_j, y(x_j))~.
\label{eq:foreu3}
\end{equation}
This employs the exact quantities ($y(x_j)$, not $y_j$), because it 
hones in on the error made in a single step, under the assumption 
that one starts from the exact solution. 
As above, using a Taylor-series expansion you can show that
the local truncation error for the Euler method is
\begin{equation}
t_{j} = \frac{1}{2} h^2 y''(\xi_j)~.
\label{eq:foreu3b}
\end{equation}
In other words, the local discretization error of this approach is $\mathcal{O}(h^2)$.
So far, things are fully analogous to the rectangle rule; the fact that there we had a first
derivative in the prefactor whereas here we have a second derivative is immaterial for our purposes.

Unfortunately, when you turn to the \textit{global} error, the analogy with the rectangle rule 
is no longer helpful. When computing the error in the composite rectangle rule, 
we took advantage of the fact that the errors in each panel were independent.
This is no longer true for the errors incurred in different steps of Euler's method (or other
single-step methods). 
Recall that when defining the local truncation error in 
Eq.~(\ref{eq:foreu3}) we assumed that we started from 
the exact solution. While this is true in the first step, it is no longer true after that:
the actual error at the $j$-th step
is therefore \textit{not} given by Eq.~(\ref{eq:foreu3b}) since 
$y(x_j)$ is actually different from $y_j$ and therefore $f(x_j, y(x_j))$ is also different from
$f(x_j, y_j)$, as a result of the errors made in previous iterations. 
Have a look at the second step in Fig.~\ref{fig:euler} to see that there the 
starting point is $y_1$, which is not the same as $y(x_1)$ 
(and therefore also $f(x_1, y_1)$ is not the same as $f(x_1, y(x_1))$).
To go beyond the local truncation error, we introduce
\begin{equation}
e_j = y(x_j) - y_j
\end{equation}
where, obviously, the total error incurred after $n$ steps, i.e., the global error, would be $e_{n-1} = {\cal E}$.
Again, this is not the same as the local truncation error, which assumes you have the correct answer
at the starting point.
It is reasonably straightforward to show that $e_j$ and $t_j$ are related as follows:
\begin{equation}
e_{j+1} = e_j + h \left [ f(x_j, y(x_j)) -  f(x_j, y_j) \right ] + t_j~.
\label{eq:foreu3c}
\end{equation}
To compute the global error ${\cal E} = e_{n-1}$, one has to carefully step through this formula, i.e.,
it is not enough to simply add up the $t_j$'s, similarly to what one does for the composite rectangle rule.
The argument that accomplishes what we need is not too difficult, but it does get a bit technical.
The end result is
\begin{equation}
|{\cal E}| \leq h \frac{M}{2L} (e^{L(b-a)} - 1)~,
\label{eq:foreu5}
\end{equation}
where $L$ is a constant and we assumed that
the second derivatives (to be found in $t_j$) are bounded by $M$.
Overall, we have arrived at an error bound for the global error in Euler's method that is
$\mathcal{O}(h)$. 
This was the same as that claimed in our quote above but, as 
seen by the complicated prefactor in Eq.~(\ref{eq:foreu5}), this error bound
cannot be trivially related to the local truncation error from Eq.~(\ref{eq:foreu3b}).
In other words, the analogy between methods that handle numerical quadrature and those 
designed for numerical integration of IVPs breaks down.

This may be a bit too abstract for some readers, so let's look at a specific example.
Consider using Euler's method to solve the following initial-value problem:
\begin{equation}
y'(x) = 1-x^2 + y(x),~~~~~~~y(0) = 0.5
\label{eq:Burden}
\end{equation}
with $x$ going from 0 to 2. 
For this problem, the constant $L$ in Eq.~(\ref{eq:foreu5}) has the value 1;
similarly, one can solve this problem analytically to see that the second derivatives
are bounded by $M = 0.5 e^2 -2$. 
Thus, Eq.~(\ref{eq:foreu5}) has turned into
\begin{equation}
|{\cal E}| \leq h \frac{1}{4} (e^{2} - 4)  (e^{2} - 1)~.
\end{equation}
For the specific case of $h=0.2$, we get the error bound $|{\cal E}| \leq 1.0826$:
this analysis doesn't tell us what the actual error will be, but whatever it is, 
we know it cannot be larger than what is mentioned in our error bound. 
Since Euler's method is very simple to code up, and this equation can be analytically
solved, it is easy to find out that the 
error that this approach leads to is $|{\cal E}| \approx 0.4397$.
As is usually the case, our \textit{a priori} error investigation 
 is too pessimistic, since the actual error incurred by 
employing Euler's method is roughly 2.5 times smaller than the value listed in the error bound.

The example discussed in the previous paragraph
is taken from the textbook by Burden, Faires, \& Burden.\cite{Burden}
This reference employs our Eq.~(\ref{eq:foreu5}), i.e.,
it does \textit{not} make Mistake \#6. However, there's
nothing stopping us from applying the mistaken error expression ourselves:  
simply adding up the local truncation errors in 
Eq.~(\ref{eq:foreu3}) leads to an estimate of the global error that is
$|{\cal E}| = h (b-a) |f''(\xi)| /2$, where $\xi$ is a point
between $a$ and $b$; this is the same as what
you would get if you translated the expression in the quote given above into our notation. 
This can be turned into an error bound using
the value of $M$ listed in the previous paragraph; plugging everything in
leads to $|{\cal E}| \leq 0.3389$. This ``error bound'' is clearly violated
by the actual error which, as we saw, is $|{\cal E}| \approx 0.4397$.
The \textit{ad hoc} summing-up of the local truncation errors led to an error
bound that is too optimistic: that's not how error bounds are supposed to work.
If you're thinking this is not such a big deal, you should reconsider: 
it is common, when using IVP solvers, to employ an error-bound formula ahead of time, to 
see which value of $h$ one should pick. If the error bound gives untrustworthy answers, one could
conceivably decide to use such a large value of $h$ that the solver becomes \textit{unstable} and therefore
gives completely worthless results!

To summarize:

{\bf Correction \#6} \textit{When solving initial-value problems,
the single-step errors are not independent from each other. Thus, simply adding
up the local truncation errors won't do. You have to see the cumulative effect the error in one panel has 
on the calculation of the next panel, and so on, until the end. 
Practically speaking, the result does end up being that the global error is one order lower than 
the local error; for example, in the forward Euler method (which has a 
local error $\mathcal{O}(h^2)$) the global error obeys an error bound that is $\mathcal{O}(h)$, but
one has to be careful with the prefactor.
}

\section{Summary and conclusion}

Above, we saw a sampling of mistaken claims in the introductory literature on computational
science. Thematically, these touched upon floating-point numbers, linear algebra, polynomial or trigonometric
interpolation, as well as stochastic integration and the solution of initial-value problems. 
Together, they constitute a good
cross section of basic themes with which every physicist should have a passing familiarity.
In each of the six cases,
we tried to provide some context that we hope helped both to explain why the misconception came about
and why it is, indeed, a mistake. 
All of these issues are very practical, which is what allowed us to bring up 
specific examples that any physics undergraduate may grasp.

It's worth highlighting that two out of the six issues (Mistakes \#3 and \#4) 
had to do with interpolation: this may be a result
of the fact that, historically, linear algebra and differential equations were considered
the two main subjects on which numerical methods focused; as a result,
topics like interpolation sometimes ``fell through the cracks,'' even in the technical literature.
We do not wish to venture a guess as to why the other four issues propagated through the 
introductory literature, 
especially since these
are handled properly in specialized volumes. 
We reiterate that our goal here is not to chastise specific authors (or, heaven forfend, an entire community),
but to promote the correct teaching of this material. That duty falls on physics instructors;
to borrow a memorable turn of phrase: 
``\textit{students cannot interrogate what they don't know.}''\cite{Judt}
In short, we hope to have made a small contribution toward the replacement of
\textit{error propagation} by \textit{error cessation}.

\begin{acknowledgments}
This work was supported in part by the Natural Sciences and Engineering Research Council (NSERC) of Canada, the Canada Foundation for Innovation (CFI), and the Early
Researcher Award (ERA) program of the Ontario Ministry of Research, Innovation and Science. 
\end{acknowledgments}

\end{document}